\documentclass[aps,prd,onecolumn,showkeys,showpacs,amssymb]{revtex4}
\usepackage{float}
\usepackage{amssymb}
\usepackage{graphicx}
\usepackage{amsmath}
\usepackage{hyperref}
\begin{document}

\title{Extreme neutron stars from Extended Theories of Gravity}

\author{Artyom V. Astashenok$^{1}$, Salvatore Capozziello$^{2,3,4}$, Sergei D. Odintsov$^{5,6,7}$}

\affiliation{$^{1}$I. Kant Baltic Federal University, Institute of Physics and Technology, Nevskogo st. 14, 236041 Kaliningrad, Russia.\\
$^2$Dipartimento di Fisica, Universita' di Napoli "Federico II and
\\$^3$Istituto Nazionale di Fisica Nucleare (INFN)  Sez. di Napoli, Complesso  Universitario  di Monte S. Angelo, Ed.
G., Via Cinthia,
9, I-80126, Napoli, Italy.\\
$^4$ Gran Sasso Science Institute (INFN),  Viale F. Crispi, 7, I-67100, L'Aquila, Italy.\\
$^5$Instituci\`{o} Catalana de Recerca i Estudis Avan\c{c}ats (ICREA), Barcelona, Spain.\\
$^6$Institut de Ciencies de l'Espai (IEEC-CSIC), Campus UAB, Torre C5-Par-2a pl, E-08193 Bellaterra, Barcelona, Spain.\\
$^7$Tomsk State Pedagogical University, 634061 Tomsk and National
Research Tomsk State University, 634050 Tomsk, Russia.}

\begin{abstract}
We discuss  neutron stars with strong magnetic mean fields in the
framework of Extended Theories of Gravity. In particular, we take
into account  models derived from $f(R)$ and $f(\cal G)$
extensions of General Relativity where functions of the Ricci
curvature invariant  $R$  and the Gauss-Bonnet invariant ${\cal
G}$ are respectively considered. Dense matter in magnetic mean
field, generated by magnetic properties of particles, is described
by   assuming a  model with three meson fields and baryons octet.
As result,  the considerable increasing of maximal mass of neutron
stars can be achieved by cubic corrections in $f(R)$ gravity. In
principle, massive  stars with $M> 4 M_{\odot}$ can be obtained.
On the other hand, stable stars with high strangeness fraction
(with central densities $\rho_{c}\sim 1.5-2.0$ GeV/fm$^{3}$) are
possible considering quadratic corrections of $f(\cal {G})$
gravity.  The magnetic field strength in the star center is of
order $6-8\times 10^{18}$ G. In general, we can say that other
branches of massive neutron stars are possible considering the
extra pressure contributions coming from gravity extensions. Such
a  feature can constitute both a probe for alternative theories
and a way out to address anomalous self-gravitating compact
systems.

\end{abstract}


 \maketitle

\section{Introduction}

The discovery of the pulsar PSR J1614-2230 \cite{Demorest} led to
several  interpretative  problems on the  physics of neutron stars
and to the possibility that the standard theory of such objects
could be revised including also the other anomalous objects that
observations  are revealing \cite{Antoniadis, Rawls}. In
particular, for a realistic description of nuclear matter, one
needs to account for the appearance of exotic particle at
densities $\sim 5-8\times 10^{14}$ g/cm$^{3}$ \cite{Nagae}.
Despite of this requirement,  the equation of state (EoS)
considerably softens ($M_{max}\sim 1.5-1.6 M_{\odot}$) (see
\cite{Glendenning,Glendenning-2,Schaffner,Vidana,Schulze}) and
therefore the maximal neutron star mass results reduced. Some
approaches  have been  recently proposed  to solve the problem
\cite{Hofmann,Rikovska, Miyatsu,
Bednarek,Weissenborn,Whittenbury,armen-1,
armen-2,armen-3,Lee,Rezzolla}, however  there is no final
agreement on the solution of the puzzle.  In particular, the
required maximal mass ($\sim 2 M_\odot$) can be obtained for
hyperon EoS from more complex model of strong interaction than
simple $\sigma\omega\omega$-model with realistic hyperon-meson
couplings. However, it is worth noticing that one cannot derive a
reliable $M$-$R$ relation for neutron stars from observations
because there are no precise radius measurements for any stars
\cite{Lattimer-2}.

On the other hand, the maximal limit of neutron star mass can increases considerably
due to strong magnetic field inside the star. The observations of
gamma-ray repeaters and anomalous X-ray pulsars may indicate
the existence of magnetic fields of the order $10^{15}$ G on the stellar
surface. In the center of the star,  the magnetic field can exceed $10^{18}$
G.

Realistic EoS with hyperons and quarks in presence of strong
magnetic field are considered in \cite{Rabhi, Rabhi-2,Ryu,Ryu-2}. The maximal mass of neutron star can
exceed, in these cases,  $3 M_\odot$ for $B_{c}\sim 3.3\times 10^{18}$ G.

Therefore it seems that the existence of neutron stars with masses
exceeding considerably two solar mass (without strong magnetic
field) is impossible in the framework of General Relativity (GR). Despite of this theoretical constraint, it is
interesting to note that there are several observational  indications in favor of
maximal masses  that exceed this limit. In fact,  masses of
B1957+20 and 4U 1700-377 are estimated as $\sim 2.4M_{\odot}$
\cite{Kerk, Clark} while for PSR J1748-2021B , the mass limit reaches $M \sim
2.7M_{\odot}$ \cite{Freire}.

In principle, observational data on neutron stars (mainly the mass-radius $M-R$
relation) can be used to investigate possible deviations  from
GR as probe for alternative gravity theories.

Alternatives to GR  have been developed in order to solve several shortcomings related to the ultraviolet and infrared behaviors of
the gravitational field as formulated in the Einstein theory \cite{report}. In particular, Extended Theories of Gravity could successfully address the
recently established phenomenon of the accelerated expansion of the universe.

This fact is confirmed by observations of type Ia supernovae
\cite{Perlmutter,Riess1,Riess2}, by microwave
background radiation  anisotropies \cite{Spergel}, by cosmic
shear through gravitational weak leasing surveys \cite{Schmidt}
and by data coming  from Lyman alpha forest absorption lines \cite{McDonald}.
This acceleration should occur  thanks  to the so called {\it dark energy}, a cosmic fluid  with
negative pressure according to the standard approach. In the framework of GR,  the
simplest explanation is given by the $\Lambda $CDM model where the Einstein
Cosmological Constant and the dark matter supply almost the 95$\%$ of the cosmic budget. However, up to now, no final answer to the question of what are  the fundamental constituents of such dark ingredients has been definitely found.

Alternative approaches to the resolution of dark side puzzle (both dark matter and dark energy)
require to modify  gravity (for details see \cite{report,Odintsov-3, Odintsov-4,Capozziello_book,Capozziello3,joice} and references therein). Main advantages of such
approaches are the possible unification of dark energy and early-time
inflation \cite{Odintsov1} and the explanation of cosmic structures without new exotic material ingredients \cite{annalen}. In principle,  these theories can satisfactorily
reproduce the data of astronomical observations and therefore one
cannot distinguish if the dark side issues can be addressed by geometry (l.h.s. of field equations) or by matter fluids (r.h.s. of field equations).

Addressing  exotic compact objects by modified gravity   could
give, in principle, new signatures in favor (or in disfavor) of
possible extensions of GR. Some models of $f(R)$ gravity can be
considered as unreliable because stable star configurations do not always
exist
\cite{Briscese,Abdalla,Bamba,Kobayashi-Maeda,Capozziello2011,Babichev2010,Nojiri5, arbuzova}.
However the existence of stable star configurations can be
achieved  in certain cases due to the so-called {\it Chameleon
Mechanism} \cite{Justin, Upadhye-Hu} or may depend on the  chosen
EoS. Furthermore,  strong gravitational regimes could be
considered if one assume GR as the weak field limit of some more
complicated effective gravitational theory \cite{Dimitri-rev}. In
this sense, higher-order corrections and extensions could be
"detected" as the mechanism capable of producing anomalous neutron
stars. In other words, the interior of a self-gravitating compact
object could behave, in broad sense,  like the early  inflationary
universe where strong gravitational fields were present.

As discussed in \cite{EKSI}, the proposition of validity of
GR  as the only theory capable of describing  neutron stars is  rather an
extrapolation because the strength of gravity within a star is
orders and orders of magnitude larger than the gravitational strengths probed
in the Solar System weak field limit tests.

Neutron star models with quadratic corrections like  $f(R)=R+\alpha R^2$
gravity are considered in \cite{Arapoglu,Alavirad,Astashenok,Astashenok-2,Kokkotas,Kokkotas-2}.  If also  magnetic fields are assumed,
more realistic models can be considered \cite{Cheoun,Astashenok-3}). In the case of these quadratic corrections,
the possible increasing (or decreasing) of maximal star mass in
comparison with GR is negligible for realistic values of the parameter $\alpha$.
Realistic models  addressing the observed models can be achieved by considering also cubic corrections in the Ricci scalar $R$ \cite{Astashenok, Astashenok-2}.

In this paper, we consider the possibility of existence of neutron
stars with extreme properties  in the framework of Extended Theories of Gravity.
In particular we will take into account generalizations of the Einstein gravity containing
functions of the Gauss-Bonnet invariant $f(\cal G)$ and we will compare them with $f(R)$
gravity.  In fact,  involving the Gauss-Bonnet invariant into the gravitational action is  a useful approach to cure shortcomings related to $f(R)$ as, for example,   the presence of ghosts. Furthermore, some functions of the topological invariant   $\cal G$ are  related to  conserved quantities (see also \cite{felix}). This fact seems to contribute to make stable  self-gravitating systems, as we will see below.

The paper is organized as follows. In Section II, we
derive  the field equations for $f(\cal G)$ and $f(R)$ gravity and the
corresponding modified Tolman-Oppenheimer-Volkoff (TOV)
equations.  In Section III is devoted to the discussion of relativistic mean field theory for dense matter. Here we show how electromagnetic properties of particles give rise  to the magnetic mean field of the neutron star and how such a field contributes to the EoS.

In Section IV, modified TOV equations are numerically solved for
realistic EoS for matter in strong magnetic field. In particular, we investigate
the cubic corrections for  $f(R)$ gravity and quadratic corrections for  $f(\cal G)$ gravity. The results are compared to GR.
Conclusions and outlook are given in  Section V.

\section{Tolman-Oppenheimer-Volkoff equations in Extended Gravity}
Let us consider now corrections to the GR where higher order
curvature invariants are added to the standard Hilbert-Einstein
action. As  discussed above, such an approach is useful to address
several issues related to  dark components.  In our case, we will
show that such corrections results as further effective pressure
in the TOV equations. We will start with dealing with $f(\cal G)$
gravity and compare results with $f(R)$ gravity.

 The  action for $R+f(\cal G)$ gravity is
\begin{equation}
\label{action}
S=\int d^4x\sqrt{-g}\left[\frac{R}{2\kappa^2}+f(\cal G)\right]+S_m\,
\end{equation}
As standard in the Einstein gravity, by working with the
commutative connections in a Riemann spacetime, $R$ is the Ricci
scalar, and the  function $f(\cal G)$ corresponds to  a generic
globally differentiable function of the Gauss-Bonnet topological
invariant ${\cal G}$. We add also the matter action $S_m$ which
induces the energy momentum tensor $T_{\mu\nu}$. We assume
$\kappa^2=8\pi G/c^4$, where $G$ is the standard  Newtonian
gravitational coupling. In metric formalism and by taking the
metric as the dynamical variable of the model, the field equations
are \cite{Nojiri2005} (see also refs. \cite{all})
\begin{equation}\label{eom}
R_{\mu\nu}-\frac{1}{2}Rg_{\mu\nu}+8\Big[R_{\mu\rho\nu\sigma}+R_{\rho\nu}g_{\sigma\mu}
-R_{\rho\sigma}g_{\nu\mu}-R_{\mu\nu}g_{\sigma\rho}+R_{\mu\sigma}g_{\nu\rho}+
\end{equation}
$$
+\frac{R}{2}\left(g_{\mu\nu}g_{\sigma\rho}-g_{\mu\sigma}g_{\nu\rho}\right)\Big]\nabla^{\rho}\nabla^{\sigma}f_{\cal G}+\left(f_{\cal G}{\cal G}-f\right)g_{\mu\nu}=\kappa^2
T_{\mu\nu},
$$
where $f_{\cal G}=d f({\cal G}) / d{ \cal G}$ and the Gauss-Bonnet  invariant is
defined as
\begin{equation}
{\cal G}=R^2-4R_{\mu\nu}R^{\mu\nu}+R_{\mu\nu\lambda\sigma}R^{\mu\nu\lambda\sigma}\,,
\end{equation}
where $R_{\mu\nu}$ and $R_{\mu\nu\lambda\sigma}$ are the
Ricci  and Riemann tensors, respectively. We adopted the
signature for the Riemannian metric as $(-+++)$. It is
$\nabla_{\mu}V_{\nu}=\partial_{\mu}V_{\nu}-\Gamma_{\mu\nu}^{\lambda}V_{\lambda}$
and
$R^{\sigma}_{\;\mu\nu\rho}=\partial_{\nu}\Gamma^{\sigma}_{\mu\rho}-\partial_{\rho}
\Gamma^{\sigma}_{\mu\nu}+\Gamma^{\omega}_{\mu\rho}\Gamma^{\sigma}_{\omega\nu}
-\Gamma^{\omega}_{\mu\nu}\Gamma^{\sigma}_{\omega\rho}$ for the
covariant  derivative and the Riemann tensor, respectively.
\par
Let us suppose now  that the metric is spherically symmetric  with coordinates $x^{\mu}=(ct,r,\theta,\varphi)$ in the following form:
\begin{equation}
ds^2=-c^2
e^{2\phi}dt^2+e^{2\lambda}dr^2+r^2(d\theta^2+\sin^2\theta
d\varphi^2)\label{g}.
\end{equation}
We assume that the interior of star is filled with a perfect fluid with
 energy-momentum tensor of the form
$T_{\mu\nu}=\mbox{diag}(e^{2\phi}\rho c^{2}, e^{2\lambda}p, r^2p,
r^{2}\sin^{2}\theta p)$,  where $\rho$ is the matter density and
$p$  the pressure. The $tt$ and $rr$ components of field
equations (\ref{eom}) read:
\begin{equation}
-\frac{1}{r^2}(2r\lambda'+e^{2\lambda}-1)+8e^{-2\lambda}
\Big[f_{\cal GG}({\cal G}''-2\lambda'{\cal G}')+f_{\cal GGG}({\cal G'})^2\Big]
\Big[\frac{1-e^{2\lambda}}{r^2}-2(\phi''+\phi'^2)\Big]+(f_{\cal G}{\cal G}-f)e^{2\lambda}=-\kappa^2\rho
e^{2\lambda}c^{2}\label{tt}.
\end{equation}

\begin{equation}
\frac{1}{r^2}(2r\phi'-e^{2\lambda}+1)+(f_{\cal G}{\cal G}-f)e^{2\lambda}=\kappa^2
p e^{2\lambda}\label{rr}.
\end{equation}
The trace of (\ref{eom}) gives
\begin{equation}\label{T}
R+8G_{\rho\sigma}\nabla^{\rho}\nabla^{\sigma}f_{\cal G}-4(f_{\cal G}{\cal G}-f)=\kappa^2(\rho
c^2-3p).
\end{equation}
The hydrostatic continuity equation follows from the contracted Bianchi identities
$\nabla_{\mu}T_{\nu}^{\mu}$ for $\nu=r$. It is
\begin{equation}
\frac{dp}{dr}=-(p+\rho c^2)\phi'\label{p}.
\end{equation}
The continuty equation is identically satisfied  for $\nu=t$. Note
that if $f({\cal G})={\cal G}$, Eqs.(\ref{tt})  and (\ref{rr}) reduce trivially to the GR field
equations being the linear $\cal G$ a conserved topological invariant resulting null when integrated in a 4D-spacetime.

Let us now replace   the metric function with the
following expression in terms of the gravitational mass function
$M=M(r)$:
\begin{equation}
e^{-2\lambda}=1-\frac{2GM}{c^2 r}\Longrightarrow \frac{G dM}{c^2
dr}=\frac{1}{2}\left[1-e^{-2\lambda}(1-2r\lambda')\right]\label{dM}.
\end{equation}
The aim is to rewrite Eqs. (\ref{tt}) and  (\ref{rr}) in terms of
${\displaystyle \frac{dp}{dr},\frac{dM}{dr}}$ and $\rho$ in a dimensionless form. For
this purpose, we introduce the following set of  dimensionless
parameters
\begin{equation}
M\to m M_{\odot},\ \ r\to r_{g} r,\ \ \rho\to \frac{\rho
M_{\odot}}{r_{g}^3},\ \ p\to \frac{ p M_{\odot} c^2}{r_{g}^3},\ \
G\to\frac{G}{r_g^4}\,.
\end{equation}

Here $r_{g}=GM_{\odot}/{c^2}$ corresponds to  one half of the
gravitational radius of the Sun. We also introduce the
dimensionless function $f\rightarrow r_{g}^{-2}f$.  From now on,
we will use  dimensionless parameters. The continuity equation
becomes  the following
\begin{equation}
\frac{dp}{dr}=-(p+\rho)\phi'\label{p2}.
\end{equation}
Eq. (\ref{dM}) becomes
\begin{equation}
\frac{d\lambda}{dr}=\frac{m}{r}\left(\frac{1-\frac{r}{m}\frac{dm}{dr}}{{2m}-{r}}\right)\label{dm}.
\end{equation}
Using Eqs. (\ref{p2}) and (\ref{dm}) in Eq. (\ref{rr}),  we obtain
\begin{equation}
\frac{2}{r^2}\left(\frac{r-2m}{p+\rho}\right)\frac{dp}{dr}+\frac{2m}{r^2}+8\pi
p-({\cal G}f_{\cal G}-f)=0\label{eq1}.
\end{equation}
For the ($tt$) equation, we get
\begin{equation}\label{eq2}
\frac{2}{r^2}\frac{dm}{dr}-8\left(1-\frac{2m}{r}\right)^2\left[f_{\cal GG}\left({\cal G}''-\frac{2m}{r}\left(\frac{1-r\frac{dm}{dr}}{{2m}-{r}}\right){\cal G}'\right)+r_g^2f_{\cal GGG}{\cal G}'^2\right]\times
\end{equation}
$$
\times\left[-\left(\frac{2m/r^3}{1-\frac{2m}{r}}\right)+2\frac{d}{dr}\left(\frac{\frac{dp}{dr}}{p+\rho
}\right)-2\left(\frac{\frac{dp}{dr}}{p+\rho}\right)^2\right]-(f_{\cal G}{\cal G}-f)=8\pi\rho.
$$
Our aim here is to solve the system of the differential Eqs.
(\ref{T}), (\ref{eq1}), (\ref{eq2}) numerically. For solving, we
use the perturbative approach described in \cite{Arapoglu} for
$f(R)$ gravity (for independent derivation of TOV equations in
$F(\cal G)$ gravity, see \cite{Momeni}). In the framework  of
perturbative approach, terms containing $f(\cal G)$ and its
derivatives are assumed to be of first order in the small
parameter $\alpha$ (i.e. $f({\cal G})\sim \alpha h({\cal G})$), so
all such terms should be evaluated at ${\mathcal O}(\alpha)$
order. Therefore one needs to calculate the Gauss-Bonnet invariant
at zero order. We have, for ${\cal G}$, the following equation
\begin{equation}
-e^{4\lambda}{\cal G}=8\frac{\left(\phi''+\phi'^2-\lambda'\phi'\right)\left(e^{2\lambda}-1\right)+2\phi'\lambda'}{r^2}\,.
\end{equation}
At zero order, the $\lambda$ and $\phi$ functions are determined by the
equations which follow from the standard TOV equations in GR, that is
\begin{equation}
\lambda'^{(0)}=\frac{1}{r}\frac{m^{(0)}-{4\pi
r^3\rho^{(0)}}}{{2m^{(0)}}-{r}}\label{TOV-1},
\end{equation}
\begin{equation}
\phi'^{(0)}=\frac{1}{r}\frac{m^{(0)}+4\pi
r^3p^{(0)}}{r-2m^{(0)}}\label{TOV-2}.
\end{equation}
In order to calculate second derivatives for $\lambda$ and $\phi$ at zero
order, one needs  the  TOV equations again. Finally we have  for
${\cal G}^{(0)}$:
\begin{equation}
{\cal G}^{(0)}={\frac { 48\ m^{(0)2}}{{r}^{6}}}-{\frac {128\pi m^{(0)}
\rho^{(0)}}{{r}^{3}}}-256{\pi}^{2}\rho^{(0)}p^{(0)}.
\end{equation}
At first order in $\alpha$, one obtains the following system
\begin{equation}\label{eq2-p}
\frac{dm}{dr}=4\pi\rho r^2+4\alpha
r^2\left[h^{(0)}_{\cal GG}\left({\cal G}''^{(0)}+\frac{2(m^{(0)}/r-4\pi
r^2\rho^{(0)})}{r-2m^{(0)}}
{\cal G}'^{(0)}\right)+h^{(0)}_{\cal GGG}{\cal G}'^{(0)2}\right]\times
\end{equation}
$$
\times
\left[\frac{2m^{(0)}}{r^3}-\frac{2m^{(0)2}}{r^4}-8\pi(\rho^{(0)}+p^{(0)})+\frac{8\pi
 m(\rho^{(0)}+3p^{(0)})}{r}-32\pi^2 r^2 \rho p\right]+\frac{1}{2}
\alpha r^2 (h^{(0)}_{\cal G}{\cal G}^{(0)}-h^{(0)}),
$$
\begin{equation}\label{eq1-p}
\left(\frac{r-2m}{p+\rho}\right)\frac{dp}{dr}=-\frac{m}{r}-4\pi p
r^2+\frac{1}{2}\alpha r^{2}(h^{(0)}_{\cal G}{\cal G}^{(0)}-h^{(0)})\,.
\end{equation}
Such modified TOV equations can be compared to the corresponding equations coming from $f(R)$ corrections.
In this case, the action is modified as

\begin{equation}
\label{actionR}
S=\int d^4x\sqrt{-g}\left[\frac{R}{2\kappa^2}+f(R)\right]+S_m\,.
\end{equation}

As before, one can assume a perturbative approach considering a perturbation parameter $\alpha$, that is  $f(R)=\alpha h(R)$.  The resulting modified  TOV system
results  (see \cite{Astashenok} for details)
\begin{equation}\label{PTOV-1}
\frac{dm}{dr}=4\pi\rho r^2-\alpha r^{2}\left[4\pi
\rho^{(0)}h^{(0)}_{R}-\frac{1}{4}\left(h^{(0)}_{R}R^{(0)}-h^{(0)}\right)\right]+
\end{equation}
$$
+\frac{1}{2}\alpha\left[h^{(0)}_{RR}\left(R'^{(0)}\left(2r-3m^{(0)}-
4\pi\rho^{(0)}r^{3}\right)+r(r-2m^{(0)})R''^{(0)}\right)+r(r-2m^{(0)})h^{(0)}_{RRR}R'^{(0)2}\right],
$$
\begin{equation}\label{PTOV-2}
\left(\frac{r-2m}{p+\rho}\right)\frac{dp}{dr}=-\frac{m}{r}-4\pi pr^2+\alpha
r^2\left[4\pi
p^{(0)}h^{(0)}_{R}+\frac{1}{4}\left(h^{(0)}_{R}R^{(0)}-h^{(0)}\right)\right]+
\alpha \left(r-\frac{3}{2}m^{(0)}+2\pi
p^{(0)}r^{3}\right)h^{(0)}_{RR}R'^{(0)}.
\end{equation}
The Ricci scalar at zero order is  $R^{(0)}=8\pi
(\rho^{(0)}-3p^{(0)})$.

For the numerical integration of Eqs. (\ref{eq2-p}), (\ref{eq1-p}) and Eqs.
(\ref{PTOV-1}),  (\ref{PTOV-2}), one needs to add  an EoS of the form
$p=p(\rho)$.

\section{Relativistic mean field theory for dense matter}

A realistic EoS can be considered assuming   dense matter in
presence of a strong magnetic field generated by the electromagnetic properties of particles.
In our perturbative approach, we do not need axially symmetric solutions for metric in order to consider a  magnetic field coupled to the star angular momentum. We are assuming here how the TOV equations are modified by the geometric terms coming from extended gravity models and how the effective magnetic field, derived from the mean quantum properties of particles,  contributes to the dynamics of the self-gravitating system.
Furthermore, it is worth stressing another important point. We are developing our considerations in the Jordan frame where matter is minimally coupled with respect to the geometry. In this case, we can easily control the relativistic  EoS that we are adopting and directly confront it with the GR counterpart (see \cite{ruben} for further details). This is not the case if we were adopting the Einstein frame. In that case, matter results non-minimally coupled with geometry and the meaning of EoS parameters could result of difficult interpretation.

As a first step, let us briefly describe the relativistic mean field theory for
dense matter. We follow the same treatment of magnetic field for
magnetic neutron stars in GR (see \cite{Rabhi,Cheoun-2,Cheoun-3,Lattimer-3}) since  the addition of
purely gravitational (geometric) terms does not change the electromagnetic
dynamics of  system. One considers the nuclear matter
consisting of baryon octet ($b=$$p$, $n$, $\Lambda$,
$\Sigma^{0,\pm}$, $\Xi^{0,-}$) interacting with magnetic field (in
general case) and scalar $\sigma$, isoscalar-vector $\omega_\mu$
and isovector-vector $\rho$ meson fields and leptons ($l=$$e^{-}$,
$\mu^{-}$). The relativistic Lagrangian is \cite{Typel}
\begin{equation}
\mathcal{L}=\sum_{b}\bar{\psi}_{b}\left[\gamma_{\mu}(i\partial^{\mu}-q_{b}A^{\mu}-g_{\omega
b}\omega^{\mu}-\frac{1}{2}g_{\rho
b}{\tau}\cdot{\rho}^{\mu})-(m_{b}-g_{\sigma
b}\sigma)\right]\psi_{b}+\sum_{l}\bar{\psi}_{l}\left[\gamma_{\mu}(i\partial^{\mu}-q_{l}A^{\mu})-m_{l}\right] \psi_{l}
\end{equation}
$$
+\frac{1}{2}\left[-(\partial_{\mu}\sigma)^{2}-m^{2}_{\sigma}\sigma^{2}\right]-V(\sigma)+\frac{1}{4}F_{\mu\nu}F^{\mu\nu}+\frac{1}{2}m^{2}_{\omega}\omega^{2}+
\frac{1}{4}\omega_{\mu\nu}\omega^{\mu\nu}+
\frac{1}{4}{\rho}_{\mu\nu}{\rho}^{\mu\nu}+\frac{1}{2}m^{2}_{\rho}{\rho}_{\mu}^{2}.
$$
Here
$\omega_{\mu\nu}=\partial_{\mu}\omega_{\nu}-\partial_{\nu}\omega_{\mu}$,
$\rho_{\mu\nu}=\partial_{\mu}\rho_{\nu}-\partial_{\nu}\rho_{\mu}$,
$F_{\mu\nu}=\partial_{\mu}A_{\nu}-\partial_{\nu}A_{\mu}$ are the
mesonic and electromagnetic field strength tensors. For
simplicity, one neglects the anomalous magnetic momenta of
particles and consider frozen-field configurations of
electromagnetic field. For density-dependent couplings
$g_{b\sigma}$, $g_{b\omega}$ and $g_{b\rho}$, we use the
parameterization by \cite{Typel}. For expectation values of meson
fields, one can obtain the following equations of motion
\begin{equation}\label{0}
m^{2}_{\sigma}\sigma=\sum_{b}g_{\sigma b} n_{b}^{s},\quad
m^{2}_{\omega}\omega_{0}=\sum_{b}g_{\omega b} n_{b},\quad
m^{2}_{\omega}\rho_{03}=\sum_{b}g_{\rho b} n_{b}.
\end{equation}
Here the scalar and vector baryon number densities are
$n^{s}_{b}$ and $n_{b}$ correspondingly. These quantities are defined as \cite{Lattimer}
\begin{equation}
n^{s}_{b}=\frac{m_{b}^{*2}}{2\pi^2}\left(E^{b}_{f}k^{b}_{f}-
m_{b}^{*2}\ln\left|\frac{k^{b}_{f}+E^{b}_{f}}{m_{b}^{*}}\right|\right),\quad n_{b}=\frac{1}{3\pi^2} k^{b 3}_{f}
\end{equation}
for neutral baryons and
\begin{equation}
n^{s}_{b}=\frac{|q_{b}|B m_{b}^{*}}{2\pi^2}\sum_{\nu}
g_{\nu}\ln\left|\frac{k^{b}_{f,\nu}+E^{b}_{f}}{\sqrt{m_{b}^{*2}}+2\nu|q_{b}|B}\right|, \quad n_{b}=\frac{|q_{b}|B}{2\pi^2} \sum_{\nu} g_{\nu}
k^{b}_{f,\nu}
\end{equation}
for charged baryons (or leptons). Here $m_{b}^{*}=m_{b}-g_{\sigma
b}\sigma$ is the so called ``effective'' mass for baryons, $E_{f}^{b}$,  $k_{f, \nu}^{b}$ are the Fermi energy and momentum defined as $E_{f}^{b}=(k_{f}^{2}+m^{*2}_{b}+2\nu |q_{b}| B)^{1/2}$ for charged particles and
$E_{f}^{b}=(k_{f}^{2}+m^{*2}_{b})^{1/2}$ for neutral particles. The
summation over $\nu=n+1/2-sgn(q)s/2$ ends at value $\nu_{max}$ at which the
square of Fermi momenta is still positive.  A summary of nucleon-meson couplings and scalar field parameters for some models in literature is given in Table I.

For hyperon-meson couplings one can assume  fixed
fractions of nucleon-meson couplings, i.e. $g_{iH}=x_{iH}g_{iN}$,
where $x_{\sigma H}=x_{\rho H}=0.600$, $x_{\omega H}=0.653$,
$x_{\rho H}=0.6$ (see \cite{Rabhi}).

\begin{table}
\label{Table1}
\begin{centering}
\begin{tabular}{|c|c|c|c|c|c|c|}
  \hline
   & $n_{s}$  & $g_{\sigma N}/m_{\sigma}$ & $g_{\omega N}/m_{\omega}$ & $g_{\rho N}/m_{\rho}$ &  &  \\
  Model & (fm$^{-3}$)  & (fm) & (fm) & (fm) & b & c \\
  \hline
  TW & 0.153 & 3.84901 & 3.34919 & 1.89354 & 0 & 0 \\
  GM1 & 0.153 & 3.434 & 2.674 &  2.100 & 0.002947 & -0.001070 \\
  GM2 & 0.153 & 3.025 & 2.195 &  2.189 & 0.003487 & 0.01328 \\
  GM3 & 0.153 & 3.151 & 2.195 &  2.189 & 0.008659 & -0.002421 \\
  \hline
\end{tabular}
\caption{The nucleon-meson couplings and parameters of scalar
field potential for some models \cite{Glendenning,Typel}. The
nuclear saturation density $n_{s}$ is also given.}
\end{centering}
\end{table}

Baryon number conservation, charge neutrality and $\beta$-equilibrium conditions allow to obtain the EoS. Matter energy density is defined as
\begin{equation}
\epsilon_{m}=\sum_{b_{n}} \epsilon^{n}_{b}+\sum_{b_{c}}
\epsilon^{c}_{b}+\sum_{l}
\epsilon_{l}+\frac{1}{2}m_{\sigma}^{2}\sigma^{2}+\frac{1}{2}m_{\omega}^{2}\omega^{2}+\frac{1}{2}m^{2}_{\rho}\rho^{2}_{0}+U(\sigma).
\end{equation}
Here
$$
\epsilon^{n}_{b}=\frac{1}{4\pi^2}\left[k^{b}_{f}(E^{b}_{f})^{3}-\frac{1}{2}m_{b}^{*}\left(m_{b}^{*}k^{b}_{f}E^{b}_{f}
+m_{b}^{*3}\ln\left|\frac{k^{b}_{f}+E^{b}_{f}}{m_{b}^{*}}\right|\right)\right].
$$
is the energy density for neutral baryons and
$$
\epsilon^{c}_{b}=\frac{|q_{b}|B}{4\pi^2}\sum_{\nu}g_{\nu}\left(k^{b}_{f,\nu}E^{b}_{f}+
(m_{b}^{*2}+2\nu|q_{b}|B)\ln\left|\frac{k^{b}_{f,\nu}+E^{b}_{f}}{\sqrt{m_{b}^{*2}+2\nu|q_{b}|B}}\right|\right)
$$
is the energy density for neutral baryons. For energy density of leptons, one needs to change $m_{b}^{*}\rightarrow m_{l}$ in the last equation.

The pressure
of dense matter is given by
\begin{equation}
p=\sum_{b_{n}} p^{n}_{b}+\sum_{b_{c}} p^{c}_{b}+\sum_{l}
p_{l}-\frac{1}{2}m_{\sigma}^{2}\sigma^{2}+\frac{1}{2}m_{\omega}^{2}\omega^{2}+\frac{1}{2}m^{2}_{\rho}\rho^{2}_{0}-U(\sigma)+\Sigma^{R}_{0},
\end{equation}
where
$$
p^{n}_{b}=\frac{1}{12\pi^2}\left[k^{b}_{f3}E^{b}_{f}-\frac{3}{2}m_{b}^{*}\left(m_{b}^{*}k^{b}_{f}E^{b}_{f}
-m_{b}^{*3}\ln\left|\frac{k^{b}_{f}+E^{b}_{f}}{m_{b}^{*}}\right|\right)\right]
$$
is the pressure for neutral baryons and
$$
p^{c}_{n}=\frac{|q_{b}|B}{12\pi^2}\sum_{\nu}g_{\nu}\left(k^{b}_{f,\nu}E^{b}_{f}-
(m_{b}^{*2}+2\nu|q_{b}|B)\ln\left|\frac{k^{b}_{f,\nu}+E^{b}_{f}}{\sqrt{m_{b}^{*2}+2\nu|q_{b}|B}}\right|\right)
$$
is the pressure for charged baryons. The rearrangement of self-energy
terms is defined by
\begin{equation}
\Sigma^{R}_{0}=-\frac{\partial \ln g_{\sigma N}}{\partial
n}m^{2}_{\sigma}\sigma^{2}+\frac{\partial \ln g_{\omega
N}}{\partial n}m^{2}_{\omega}\omega^{2}_{0}+\frac{\partial \ln
g_{\rho N}}{\partial n}m^{2}_{\rho}\rho^{2}_{0}.
\end{equation}

For effective EoS, one needs also to account for  the energy density and pressure generated from the magnetic field. One obtain
\begin{equation}
\epsilon_{f}=\frac{B^2}{8\pi},\quad
p_{f}=\frac{B^2}{8\pi}.
\end{equation}
Clearly,  the magnetic field depends only on the baryon density
$n$. We use a simple parameterization (see, for example,
\cite{Rabhi, Ryu-2})
\begin{equation}
B=B_{s}+B_{0}\left(1-\exp\left(-\beta(n/n_{s})^{\gamma}\right)\right),
\end{equation}
where $B_s$ is the magnetic field on star surface ($10^{15}$ G).
For parameters $\gamma$ and $\beta$ one can realistically assume  the values
$\gamma=2$, $\beta=0.05$ (slowly varying field) and $\gamma=3$,
$\beta=0.02$ (fast varying field). The value
$B_{0}=B_{c}B_{0}^{*}$ where $B_{c}=4.414\times 10^{13}$ G
is the critical field for electrons.

\begin{table}
\label{Table1}
\begin{centering}
\begin{tabular}{|c|c|c|c|c|c|}
  \hline
  $B_{0}$,  & $\beta$, & $M_{max}/M_{\odot}$ & $R$,  & $E_{c}$, & $B_{c}$, \\
  $10^5$ & $10^{2}r_{g}^{4}$ &  & km & GeV/fm$^{3}$ & $10^{18}$ G \\
  \hline
    & 0     & 2.73 & 12.44 & 0.93 & 4.19 \\
2 & $-0.50$ & 3.54 & 12.60 & 0.79 & 3.67 \\
  & $-0.75$ & 3.92 & 12.63 & 0.78 & 3.61 \\
\hline
& 0     & 2.98 & 13.78 & 0.76 & 3.86 \\
  3 & $-0.50$ & 3.90 & 13.86 & 0.69 & 3.56 \\
& $-0.75$ & 4.16 & 13.99 & 0.65 & 3.38 \\
   \hline
& 0     & 3.15 & 14.34 & 0.68 & 3.69 \\
  4 & $-0.50$ & 4.03 & 14.47 & 0.63 & 3.45 \\
& $-0.75$ & 4.62 & 14.59 & 0.59 & 3.32 \\
   \hline
\end{tabular}
\caption{Compact star properties (maximal mass and corresponding
radius) using TW model for cubic $f(R)$ gravity for certain values
of $\alpha$ (in units of $r_{g}^4$) for fast varying magnetic
field. The magnetic field $B_{c}$ and energy density $E_{c}$ in
center are given.}
\end{centering}
\end{table}

\begin{table}
\label{Table2}
\begin{centering}
\begin{tabular}{|c|c|c|c|c|c|}
  \hline
  $B_{0}$,  & $\beta$, & $M_{max}$, & $R$,  & $E_{c}$, & $B_{c}$, \\
  $10^5$ & $10^2{r}_{g}^{4}$ & $M_{\odot}$ & km & GeV/fm$^{3}$ & $10^{18}$ G \\
  \hline
      & 0 & 2.80 & 13.99 & 0.79 & 3.49 \\
  2  & $-0.5$ & 3.21 & 13.68 & 0.86 & 3.73 \\
   \hline
  & 0 & 3.21 & 15.67 & 0.63 & 3.29 \\
   3 & $-0.5$ & 3.55 & 15.52 & 0.65 & 3.39 \\
   \hline
     & 0 & 3.53 & 17.16 & 0.52 & 3.04 \\
   4 & $-0.75$ & 4.01 & 16.93 & 0.54 & 3.13 \\
   \hline
\end{tabular}
\caption{Compact star properties using TW model for cubic $f(R)$
gravity for certain values of $\beta$ for slowly varying magnetic
field.}
\end{centering}
\end{table}

\begin{figure}
  \includegraphics[scale=1.1]{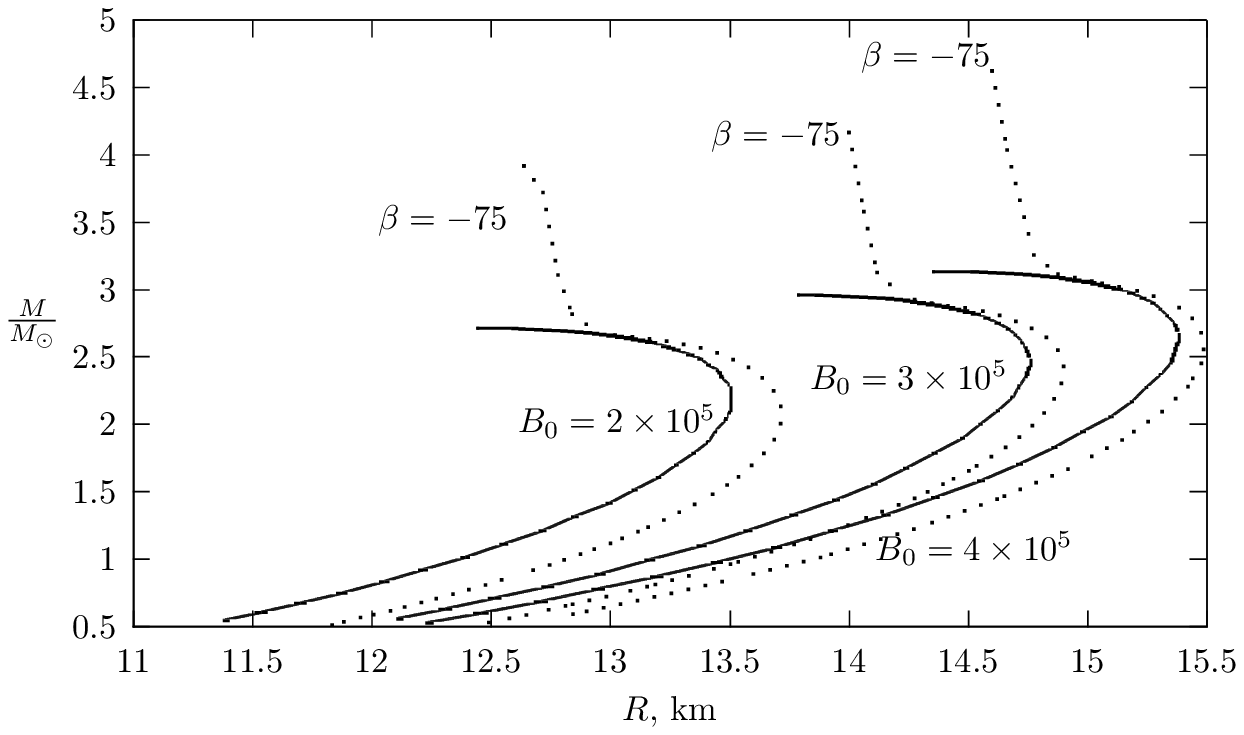}\\
  \includegraphics[scale=1.1]{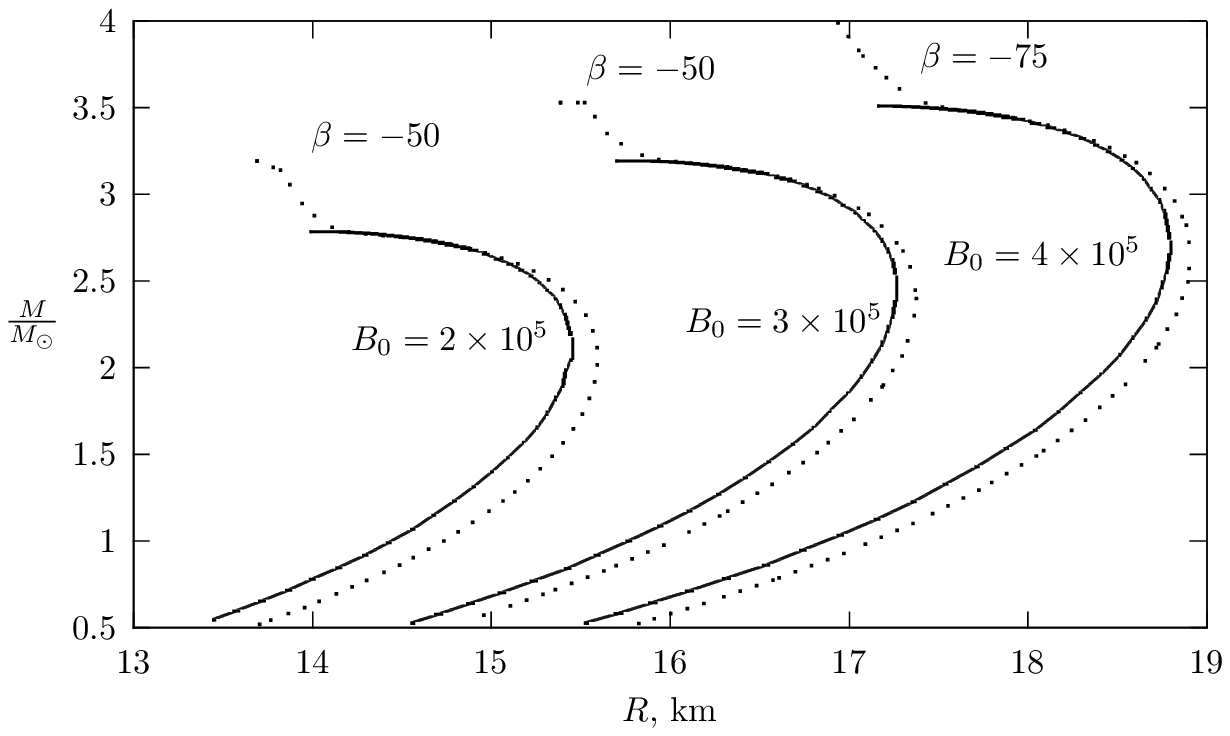}\\
  \caption{The mass-radius diagram in model $f(R)=\beta R^3$ (dotted lines) and in General Relativity (solid lines) for fast (upper panel) and slowly (lower panel) varying field.}
\end{figure}

\section{Results}
The  modified TOV equations,  equipped with the above EoS (we use
TW parametrization), can be used to construct self-consistent
models for extreme neutron stars. Here we will develop two
specific cases where cubic corrections for $f(R)$ gravity and
quadratic corrections for $f(\cal G)$ gravity are adopted.  The
general result is that stable configurations for extremely massive
neutron stars with strong magnetic fields can be achieved.

\subsection{ The case of $f(R)$ gravity.}

Although the possible existence of supermassive neutron stars is
discussed in \cite{Astashenok-3},  in the present paper this
question is considered in more detail. We consider more variants
 of magnetic field profile (various $B_{0}$). Furthermore, we
considered also the  speculative question about stars with masses
$\sim 4 M_{\odot}$ in f(R) gravity. For cubic corrections of the
form  $f(R)=\beta R^3$,  massive neutron stars with $M>3M_{\odot}$
can exist (for $\beta<0$). The small parameter $\alpha$ is now
$\alpha=\beta |R^{(0)3}_{max}|$. The parameter $R^{(0)}_{max}$
corresponds to the maximal value of the Ricci scalar. This value
strictly depends on the EoS.  We take into account  the cases
$B_{0}=2,$ $3$, $4\times 10^5$. The value of the parameter is
$\alpha<4\times 10^{-3}$ for realistic densities related to  these
cases. The properties of stars with maximal mass are given in
Tables II and III  and are compared to GR  ($\beta=0$). The
mass-radius relation is represented  in Fig. 1. It is worth
noticing  that,  although the maximal mass increases, the central
density decreases (for fast varying field). In the case of fast
varying magnetic field, the increasing of mass can exceed $1.5$
solar masses for certain values of parameter $\beta$. The maximal
possible mass is close to $4.5 M_{\odot}$ for $B_{0}=4\times 10^5$
and $\beta=-75$ (in units of $r_{g}^{6}$). The minimal value of
parameter $\beta$, at which the increasing of maximal mass is
$\sim 1 M_{\odot}$, is $\beta\sim -60$ for these values of
$B_{0}$. In the case of slowly varying field,  the minimal value
of $\beta$ for $\Delta M\sim 0.5 M_{\odot}$ is $\beta\sim -60$ for
$B_{0}=2-3\times 10^5$ and $\beta\sim -75$ for $B_{0}=3-4\times
10^5$.

The curvature in zeroth order $R^{(0)}$ is $\sim
\rho^{(0)}-3p^{(0)}$. For large densities,  in the case of magnetic
field, the value $\rho^{(0)}-3p^{(0)}$ changes sign. For example,
for fast varying magnetic field,  the change of sign corresponds
to $\rho\approx 580 $ MeV/fm$^{3}$ ($B_{0}=2\times 10^{5}$),
$\rho\approx 440 $ MeV/fm$^{3}$ ($B_{0}=3\times 10^{5}$) and
$\rho\approx 400 $ MeV/fm$^{3}$ ($B_{0}=4\times 10^{5}$). At these
densities, the $M-R$ diagram begins to  diverge from curve in GR. The
``effective'' density ($\frac{1}{4\pi r^2}\frac{dM}{dr}$) increases
and we get larger mass than in GR at the same radii.

A consideration is in order at this point.
In the cases we are dealing with,  the particle $\Sigma^{-}$ is the  hyperon that  contributes to the EoS.  In this picture, this is the only strange particle contributing to the stars of maximal mass.   For
$B_{0}=2\times 10^5$ (slowly varying field) there are also $\Lambda$
hyperons. In contrast to this, for $B_{0}=4\times 10^5$ (fast
varying field), the maximal possible mass corresponds to densities where
 hyperons do not  appear yet. It is interesting to see that this picture takes place both in
GR and $f(R)$ gravity. This means that extreme neutron stars can be achieved without exotic particles out of the Standard Model.

Some considerations are needed on the  validity of perturbative approach. For
neutron stars,  one can estimate curvature as $10^{-2}$ (in units of
$r_{g}^{-2}$) and therefore one can conclude that the term $\beta
R^{3}$ is comparable with $R$ for $\beta\sim 10^{4}$. For $\beta$,
we considered much smaller values. Of course the question arises why
this lead to substantial departure from GR? The
situation can be explained in the following way. For some EoS,  we
have the  rapid growth of mass (without substantial change of
radii) for some range of central densities. In GR,
then we have decreasing  stellar masses. But $f(R)$ terms lead to
the change of density profile and then the mass increases. In fact the,
``effective'' EoS changes so that stable stars can
exist at high central densities.

\begin{table}
\label{Table1}
\begin{centering}
\begin{tabular}{|c|c|c|c|c|c|}
  \hline
  $B_{0}$,  & $\beta$, & $M_{max}/M_{\odot}$ & $R$,  & $E_{c}$, & $B_{c}$, \\
  $10^5$ & $r_{g}^{6}$ &  & km & GeV/fm$^{3}$ & $10^{18}$ G \\
  \hline
    & 0       & 2.73 & 12.44 & 0.93    & 4.19 \\
  2 & $-0.05$ & 2.74 & 12.51 & 2.03   & 7.29   \\
    & $-0.1$  & 2.74 & 12.53 & 1.43 & 5.81   \\
\hline
    & 0       & 2.98 & 13.78 & 0.76    & 3.86 \\
  3 & $-0.05$ & 2.99 & 13.76 & 2.11   & 7.92    \\
    & $-0.1$ & 2.99 & 13.79 &  1.58   & 6.57    \\
   \hline
    & 0       & 3.15 & 14.34 & 0.68    & 3.69 \\
  4 & $-0.05$ & 3.16 & 14.36 & 2.15   & 8.75    \\
  & $-0.1$    & 3.16 & 14.42 & 1.64  & 6.92     \\
   \hline
\end{tabular}
\caption{Compact star properties using TW model for quadratic
$f(\cal G)$ gravity for certain values of $\beta$ (in units of
$r_{g}^6$) for fast varying magnetic field.}
\end{centering}
\end{table}

\begin{figure}
  \includegraphics[scale=1.1]{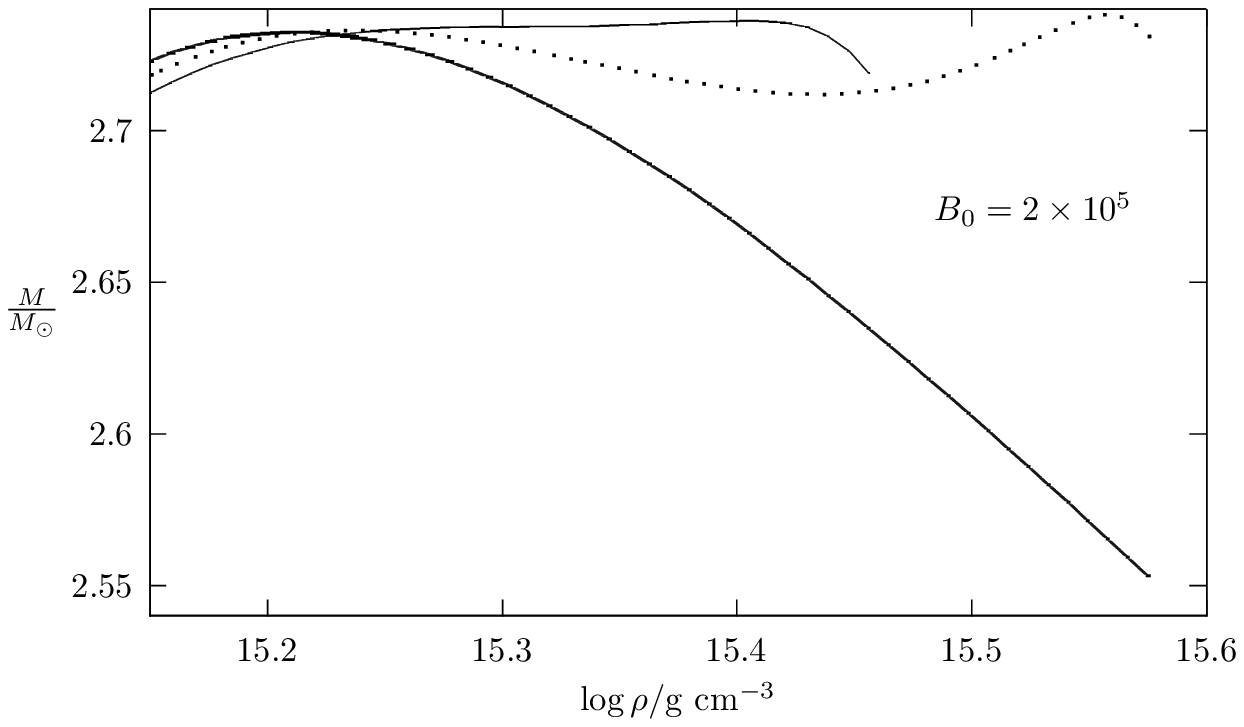}\\
  \includegraphics[scale=1.1]{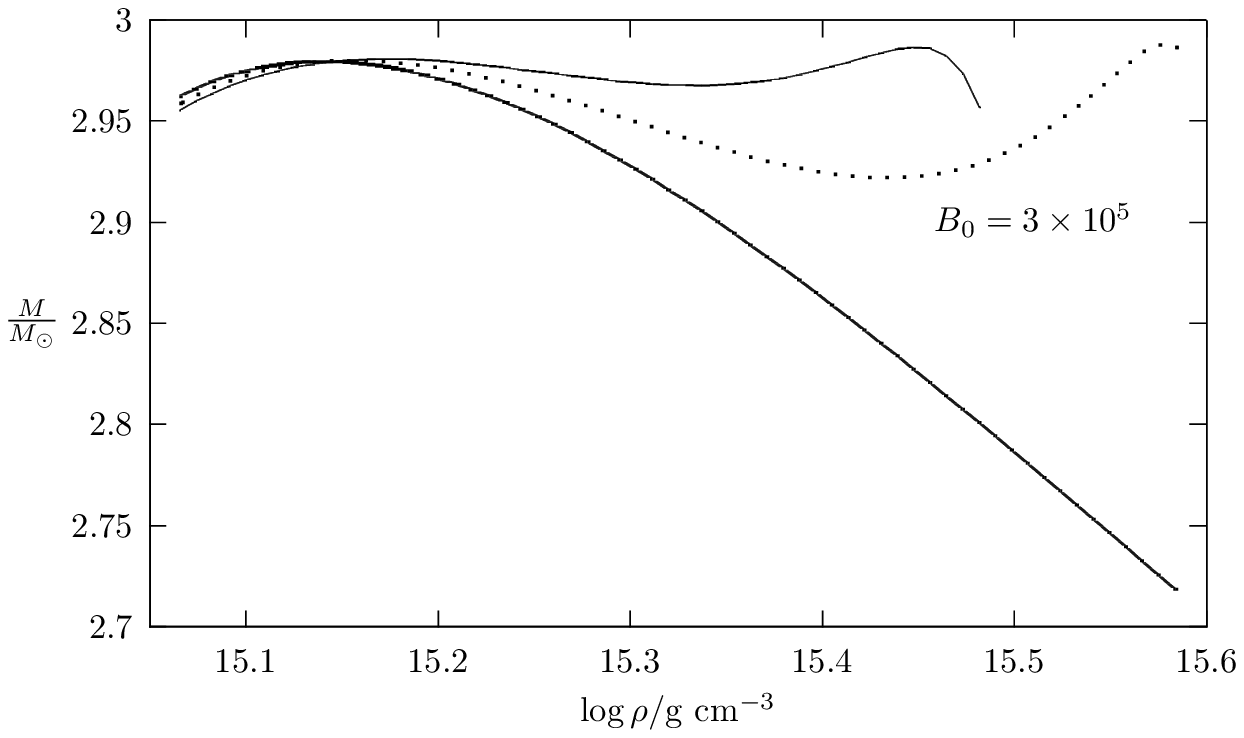}\\
  \includegraphics[scale=1.1]{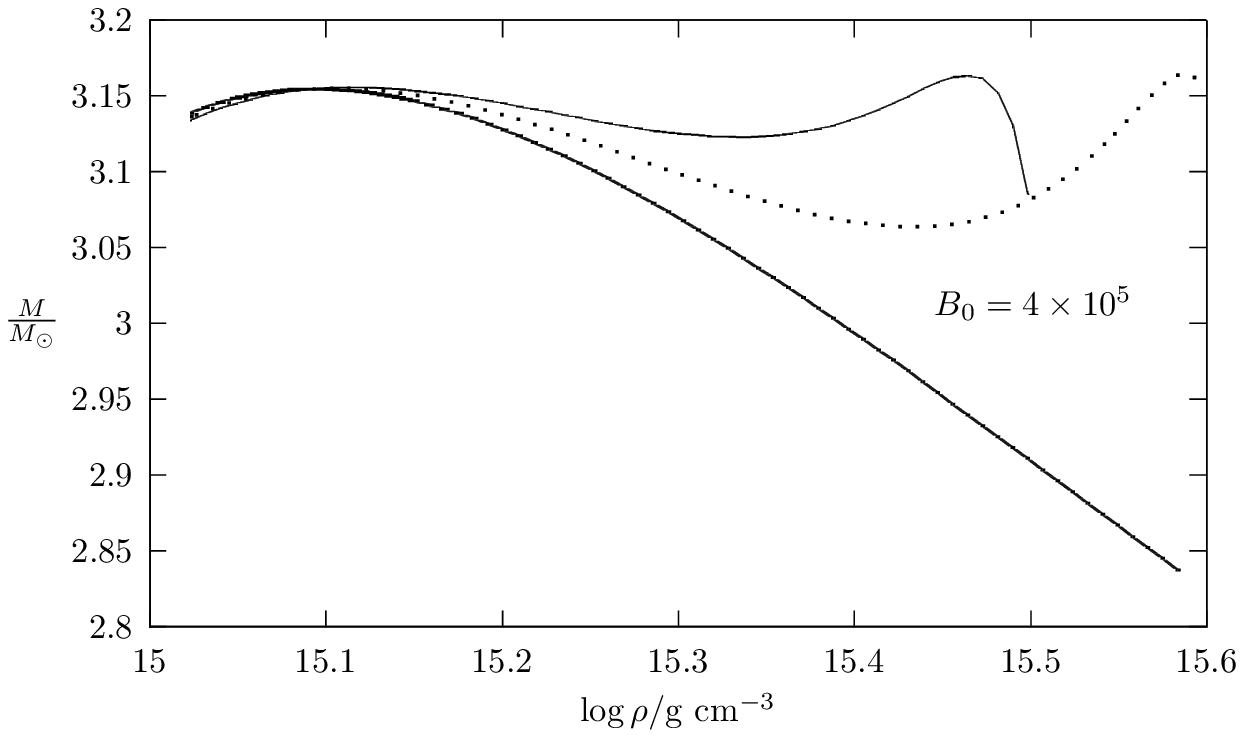}\\
  \caption{The mass-density diagram in the vicinity of maximal mass for the  model $f({\cal G})=\beta {\cal G}^2$  ($\beta=-0.05$ - dotted lines, $\beta=-0.1$ - thin
  lines)
  and for  General Relativity (thick lines) for fast varying field. There is a second branch of
  stability ($dM/d\rho>0$)
  corresponding to stars with large central densities. However, the maximal possible mass and radius  are close to the values in General Relativity.}
\end{figure}

\subsection{The case of  $f({\cal G})$ gravity}

The case of  Gauss-Bonnet quadratic corrections $f({\cal G})=\beta
{\cal G}^2$ is interesting because allows to achieve  stable
neutron star configurations with high central densities for
negative $\beta$. The situation is more clear from an intuitive
point of view in comparison with $f(R)$ gravity:  masses and
radii of stars differ from GR insignificantly because the term
$\sim G^2$ is small in comparison with $R$ term. But the character
of mass-density dependence changes: the mass begins to increase with
the increasing of central density in a narrow range of densities.
Although this increasing is small,  $dM/d\rho_{c}>0$ and
stability of star configurations is achieved at high central
densities.

Therefore the existence of stable neutron stars with extremely
magnetic fields in the center (in comparison with GR) is possible
in this model.  Although the increasing of mass is negligible, the
central energy density for stars with maximal mass is close to
$1.5-2.0$ GeV/fm$^3$ (see Fig. 2 for mass-density relation in the
case of fast varying field. For slowly varying field similar
effects take place). The cores of such stars can contain
considerable fraction of $\Sigma^{-}$ and $\Lambda$ hyperons. The
corresponding field in the center of the star can exceed
$7-8\times 10^{18}$ G (in GR, the maximal central field for these
models is only $\sim 4.2\times 10^{18}$ G, see Table IV). It is
interesting to note that similar effect takes place also in the
case of quadratic $f(R)$ gravity (see \cite{Astashenok-3}). It is
worth  noticing that, for $\beta<\sim -0.2$ and quadratic $f(\cal
G)$ corrections, there is no stable stars with extremely large
central densities.

\section{Discussion and Conclusions}

We have considered neutron star models with strong magnetic fields
in the framework of $f(R)$ and $f(\cal G)$ gravity models. We have
used a
 model with three meson fields for dense matter in strong
magnetic field and coupled the corresponding EoS to the modified
TOV equations. The reason to adopt such an approach is to
investigate the  possibility of the existence of neutron stars
with extreme features as  high central densities and  large masses
as recently pointed out by observations of some peculiar objects.

Our considerations show that considerable increasing of mass can
be achieved adopting  cubic $f(R)$ gravity corrections. Thus, the
possibility of supermassive ($M>4M_\odot$) neutron stars with
$R\sim12-15$ km in modified gravity seems, in principle,
realistic. If such stars will be explicitly observed,  this could
be considered as  a clear signature  that some self-gravitating
systems can violate   General Relativity constraints  in favor of
modified gravity.

On the other hand,  quadratic $f(\cal G)$ gravity corrections
indicate  that   another interesting effect is possible: namely
stable stars with central densities close to $\rho_{c}=1.5-2.0$
GeV/fm$^{3}$ (and therefore with high strangeness fraction) can
exist. The field strength in the center can exceed $8\times
10^{18}$ G. This limit cannot be achieved in General Relativity by
using standard observed matter.

As a general remark, we can say that the puzzles related to the
existence of extreme neutron stars could be realistically
addressed by supposing the emergence of corrections and extensions
to the General Relativity in the strong field regimes. In some
sense, the mechanism could be similar to that supposed in the
early-time inflation where higher-order curvature terms naturally
emerge into dynamics. Beside the explanation of anomalous compact
star, this approach could be considered as an independent probe
for modified gravity with respect to the analogue descriptions
invoked for dark matter and dark energy.

\acknowledgments This work is supported in part by projects
14-02-31100 (RFBR, Russia) (AVA), by MINECO (Spain), FIS2010-15640
and by MES project TSPU-139 (Russia) (SDO). SC is supported by
INFN ({\it iniziative specifiche} TEONGRAV and QGSKY).

\end{document}